\documentclass[twocolumn]{revtex4}
\usepackage{graphicx}
\newcommand{\s}{{G}}
\newcommand{\g}{{S}}

\date{\today}

\begin{document}

\title{\textbf{MicroRNA Interaction network in human:} \textit{implications of clustered microRNA in biological pathways and genetic diseases}}

\author{Sushmita Mookherjee$^1$}
\author{Mithun Sinha$^2$}
\author{Saikat Mukhopadhyay$^3$}
\author{Nitai P. Bhattacharyya$^{2,3}$}
\email[E-mail address: ]{nitaipada.bhattacharya@saha.ac.in}
\author{P. K. Mohanty$^4$}
\email[E-mail address: ]{pk.mohanty@saha.ac.in}
\affiliation{$^1$Centre for Applied Mathematics and Computational Science,
$^2$Structural Genomics Section,  $^3$Crystallography and Molecular Biology Division,$^4$Theoretical Condensed Matter Physics Division, \\Saha Institute of Nuclear Physics,1/AF Bidhan Nagar, Kolkata, 700064 India.  }
\begin{abstract}
 A novel group of small non-coding RNA, known as microRNA (miRNA) is predicted to regulate as high as 90\% of the coding genes in  human. The diversity and abundance of miRNA targets offer an enormous level of combinatorial possibilities and suggest that miRNAs and their targets form a complex regulatory network. In the present study, we analyzed $711$ miRNAs and their $34,525$ predicted targets  in the miRBase database (http://microrna.sanger.ac.uk/, \textit{version 10}), which generate a complex bipartite network having numerous numbers of genes forming the hub. Genes at the hub (total $9877$) are significantly over represented in genes with specific molecular functions, biological processes and biological pathways as revealed from the analysis using PANTHER (http://www.pantherdb.org/genes/). We further construct a   miRNA co-target network by linking every 
pair of miRNAs which co-target at least one gene. The weight of the link,  which is taken to be the number 
of co-targets of the pair of miRNAs vary widely, and we could erase several links while keeping the relevant features of the network intact. The largest connected sub-graph, thus obtained, contains 479 miRNAs. More than $75\%$ of the miRNAs deregulated in 15 different diseases 
collected from published data are found to be in
this largest sub graph. We further analyze this  sub-graph to obtain 70 small clusters containing total $330$ miRNAs of $479$. We identified the biological pathways where the co-targeted genes in the clusters are significantly over- represented in comparison to that obtained with that are not 
co-targeted by the miRNAs in the cluster. Using published data, we identified that specific clusters of miRNAs are associated with specific diseases by altering particular pathways. We propose that instead of single miRNA, clusters of miRNA that co-targets the genes are important for the regulation of miRNA in diseases.
\end{abstract}
\maketitle
\section{Introduction}

Micro-RNA (miRNA) belongs to a class of small non-coding single stranded RNA, approximately 21 nucleotides long, which negatively regulate gene expressions. Mature miRNA interacts with the $3^\prime$ untranslated regions ($3^\prime$ UTR) of the gene in human and down regulate the expression of the target either by degrading the mRNA or inhibiting the translation. In some cases, increased expressions of the target gene by miRNAs have also been reported (reviewed in [1]). Recent experiments show, at least in few specific cases, that the mature miRNA can alters the expressions of the genes by binding to the coding regions as well as $5^\prime$ UTR of genes [2-4] providing further complex regulation of the genes by miRNAs. It has been proposed on the basis of theoretical analysis that as many as 30\% of genes in the human genome may be the target of miRNA [5]. Recent, estimates predict that as large as 90\% human genes are targets of miRNA [6]. However, experimental validation of such prediction is largely lacking. Function of each region of the mature miRNA is not well defined, although, the seed region (2$^{nd}$ to 7$^{th}$ position from the $5^\prime$ end of the mature miRNA), is the most important region that interacts with $3^\prime$ UTR for regulation of the target genes. The other regions known as extended seed and delta seed regions also contribute to the target selection [7]. The diversity and abundance of miRNA targets offer an enormous level of combinatorial possibilities and suggest that miRNAs and their targets appear to form a complex regulatory network.

Functions of miRNA is known only for some, although deregulation of miRNAs has been shown in number of diseases like various types of cancers [8, reviewed in 9] cardiovascular development and heart failure [10]. Involvement of miRNAs in various diseases has been recently reviewed [11]. Studies of various diseases and normal cellular processes indicate that miRNAs are involved in immune-system [12]; stem cell renewal and development [13, 14]. Using inhibitors of different miRNAs it has also been shown that several miRNAs are involved in cell death, cell growth and proliferation [15].  The extent of modulation of the targets and their influences on the biological processes that lead to alteration of cellular phenotype varies considerably [16]. In a recent study in Caenorhabiditis elegans (C. elegans) where $83\%$ of the C. elegans miRNA (total $95$ miRNA) are mutated and effects of these
mutant genes are studied. It is shown that majority of miRNA mutations do not result into any phenotypic changes, indicating that there are redundancies of miRNAs that can target a gene. However, $10\%$ of miRNA deletion causes clear developmental and morphological defects. It has been proposed that there is significant functional redundancy among miRNAs or among gene pathways regulated by miRNAs [17]. In spite of considerable information on the involvement of miRNA in numerous biological processes and large number of human diseases, the precise information of the targets they regulate in the normal processes and diseases remains elusive.

 Role of miRNAs in signal transduction pathway has been identified in a recent work.
 By analyzing the interactions between miRNAs and a human cellular signaling network, it has been observed that miRNAs specifically target positive regulatory motifs, downstream network components and the highly connected scaffolds in a signaling network and genes whose promoter regions include a large number of putative transcription factor(TF) binding sites [18]. Combinatorial regulation of genes by TF and miRNAs provides higher complex 
programs  [19]. However, it is not known fully how miRNA-miRNA interactions regulates the expressions of the genes. Interactions of miRNA-miRNA and miRNA-TF in regulating the expressions of protein coding genes have recently been studied. Two databases namely Target Scan and PicTar have been used that cataloged $8672$ predicted targets of $138$ miRNA and $9152$ predicted targets of 178 miRNA respectively. There are overlaps ($~80\%$) in the miRNA lists and targets covered in these two databases [20]. Abilities of three miRNA namely miR-16, miR-34a, miR-106b to alter cell cycle, target expressions and apoptosis have been studied. It has been observed that miR-16 and miR-34a together block G1 to S transition greater than that obtained individually but lesser than additive. On the contrary, expression of miR-106b, which accelerates G1 to S transition, together with miR-16 or miR-34a, reduce the cells in G1 less than that obtained with miR-16 or miR-34a alone but higher that obtained with miR-106b alone. This result shows that miR-16 and miR-34a together exhibit an intermediate cellular phenotype. All these miRNAs alter specific sets of the targets. Thus different miRNA may target the different proteins in the same pathways [21]. Experimental evidence also has been provided recently to show that specific pairs of miRNAS together involves in the maintenance of embryonic stem cells [22].  It is important to know whether miRNAs interact in a combinatorial fashion, alters specific biological pathway(s) and participates in human diseases.

In the present communication, using 711 miRNA in human and their predicted 
targets we have performed a topological analysis of 
miRNA network to elucidate  how miRNA-miRNA interactions 
regulate the targets. We observe that the miRNA are clustered and that 
some clusters of miRNA 
co-target the genes in the specific pathways. Analysis of published experimental data 
describing  deregulated miRNA and mRNA in 15 diseases, support this notion.  

\section{Results and Discussions}
\subsection{Connectivity in miRNA- gene interaction network}
For the analysis of miRNA and their targets in human, we use the miRBase database, (http://microrna.sanger.ac.uk/) which has $M=711$ microRNAs (miRNAs) and 
$N=34525$ gene targets. First, a data set is created where the gene versus the set of miRNAs targeting that particular gene is listed row-wise. A schematic representation of the data set  as a  bi-partite network  is shown in 
Figure 1. The  blue and red circles  there  represent the miRNAs  and the genes respectively.
For convenience, both miRNAs and genes are given 
arbitrary identification number $m=1,2, \dots i,\dots M$  and $n= 1,2,\dots j,\dots N$ respectively. 
A line  (or link) is drawn between micro RNA $i$ and a gene $j$, 
if $i$ targets $j$. In total there are $676265$ links which connect the miRNAs and  their targets. 
From the figure it is evident that the system of miRNA-gene form a bi-partite network represented by a 
$(M\times N)$ adjacency matrix $A$ with matrix elements, 
 \begin{equation}
A_{ij} = \left\{ \begin{array}{cc} 1 & \textrm{if~miRNA~ {\it i} ~targets~ gene ~{\it j} } \cr 0&  \rm{ otherwise} \end{array}\right. .
\nonumber
\end{equation}
 
\begin{figure}
\centerline{\includegraphics[width=8cm]{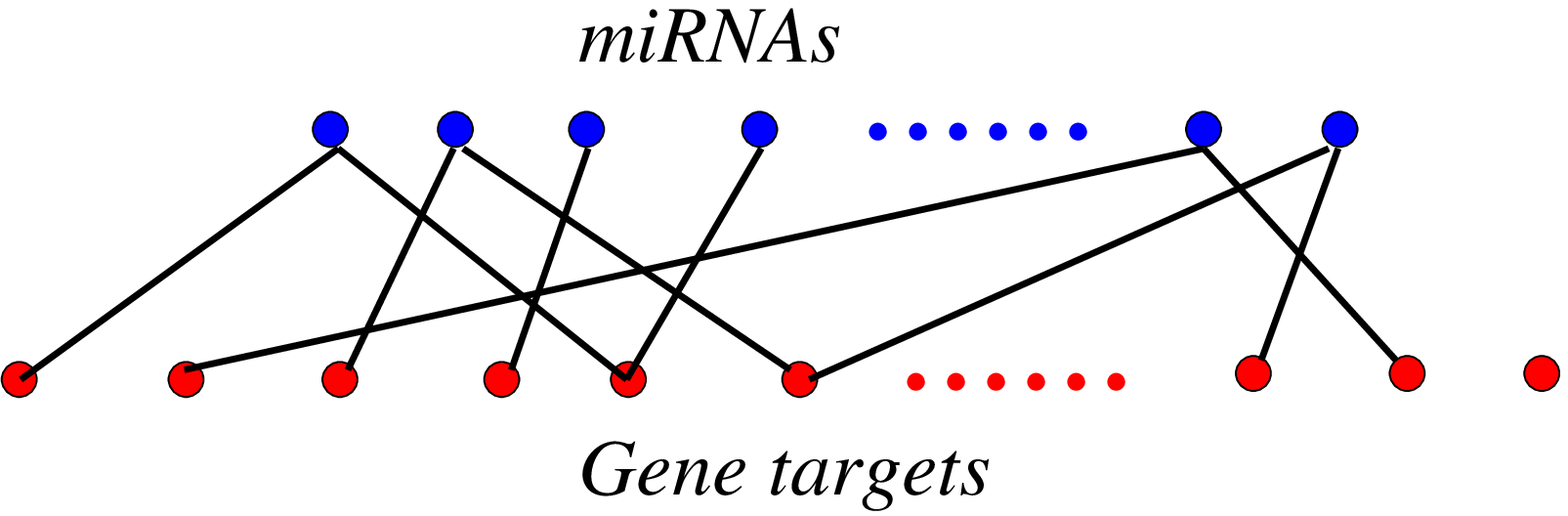}}
\textbf{Figure 1:} Figure shows a schematic representation of the bipartite connections of miRNAS and their 
targets.
\end{figure}

The network is bi-partite, because there is no link between two elements within the group of genes or miRNAs, i.e. two genes or two miRNAs are never connected. 
In Figure 2A, we have shown the distribution of targets (genes) $P_g(n)$, 
which is the  
fraction of genes (shown as circles in Figure 1) targeted by $n$ number of miRNAs. If the connections had been random (i.e.,  if $676265$ links were distributed among $711$ miRNAs randomly) we would have obtained a normal distribution (blue solid line in Figure 2A). It is clear that $P_g(n)$ varies significantly from what is expected from a random bi-partite graph.  $P_g(n)$ could be fitted to an exponential distribution (shown in the inset of Figure 2A) with a
a typical scale  $k^* = 20$. This implies that, most of the genes are targeted by only $k < 20$ miRNAs.  
Only a few genes are targeted by a large number ($k\ge 20$) of miRNAs which are termed as the target {\it hubs}. 
The degree distribution, being different from  a random normal graph,   implies that there is a high 
organizational structure in the gene-miRNA interaction network. Similar 
organizational structure  
is observed from  the distribution of miRNAs $P_m(k)$,  which is  the fraction of miRNAs that 
target $k$ number of genes.
We find  that $126$ miRNA in the miRNA hub target $\ge 1067$ genes (data not shown).

\begin{figure}
\centerline{\includegraphics[width=8cm]{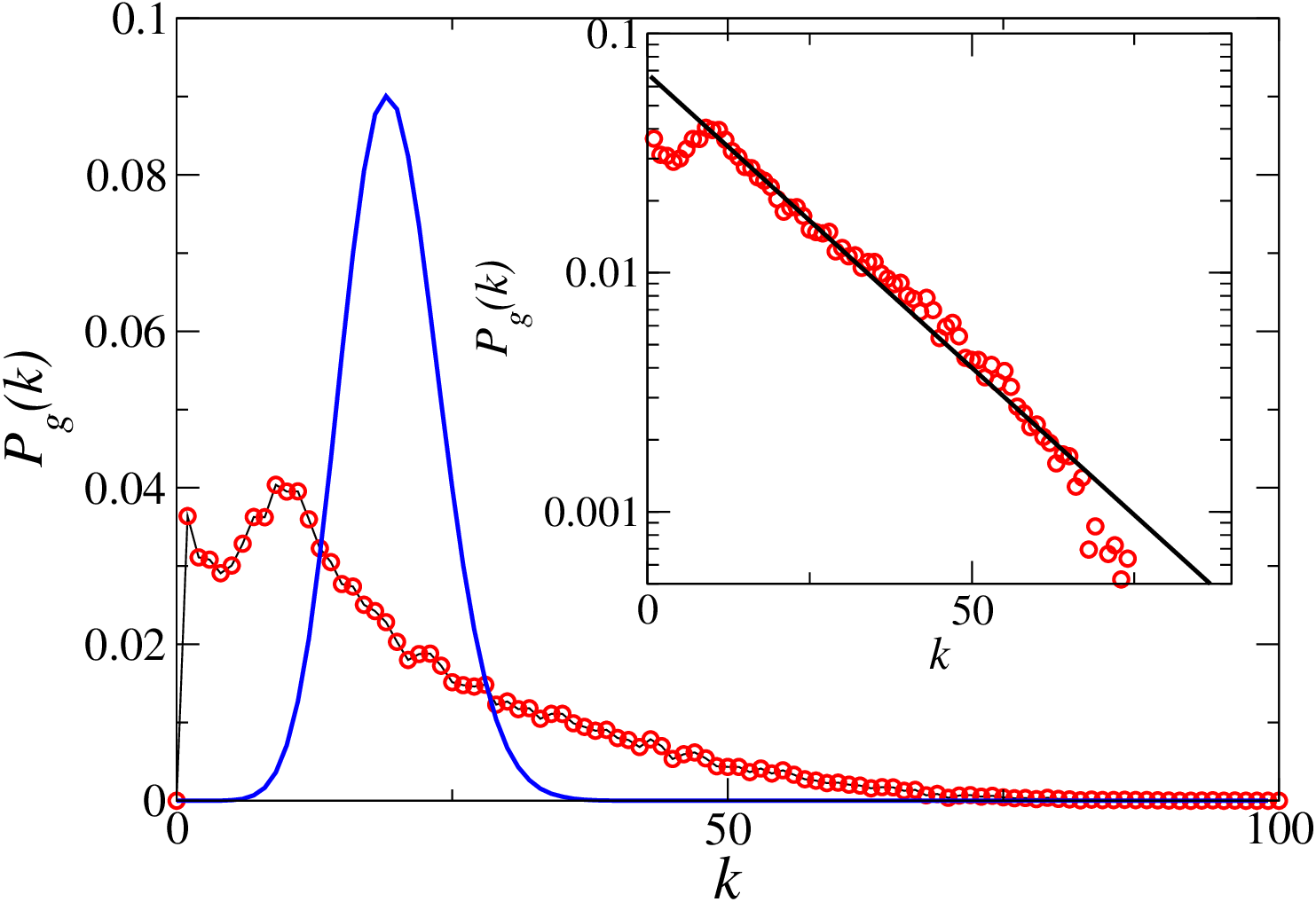}}
\textbf{Figure 2A:} Distribution of targets $P_g(k)$, which is fraction of  
genes targeted by $k$ miRNAs (red circles) is compared with a random  network (blue lines). 
The same plot is shown in the inset in semi-logarithmic scale. 
\end{figure}

Analysis of the genes in the target {\it hub} (total $9877$, Suplimentary table S1A) by 
PANTHER revealed that 
genes with molecular functions like nucleotide binding, transcription factor, receptor and hydrolase 
are enriched compared with the genes in the human genome. Genes involved in biological process like 
apoptosis, cell cycle, developmental process, nucleic acid metabolism, signal transduction etc. are  
 significantly over represented among the genes in the target hub. Genes involved in cell proliferation and differentiation, cell structure and motility and oncogenesis are also significantly decreased.  Genes in the target hub are  enriched in specific pathways like p53 pathway, angiogenesis, in two neurodegenerative disease pathways namely Alzheimer’s disease and Huntington’s disease pathway. Representative result of pathways enrichment 
is shown in the Figure 2B and the full analysis is shown in the 
Supplementary table $S1B$ to $S1D$. Results shown in Supplementary table  $S1B$ to $S1D$ and 
in the Figure 2B indicate that miRNA in combination may alter molecular functions, 
biological processes and cellular pathways. 
This result is similar to that have been obtained with lesser number of 
miRNA (about $180$) that target approximately $10000$ genes [20].

\begin{figure}
\centerline{\includegraphics[width=8cm]{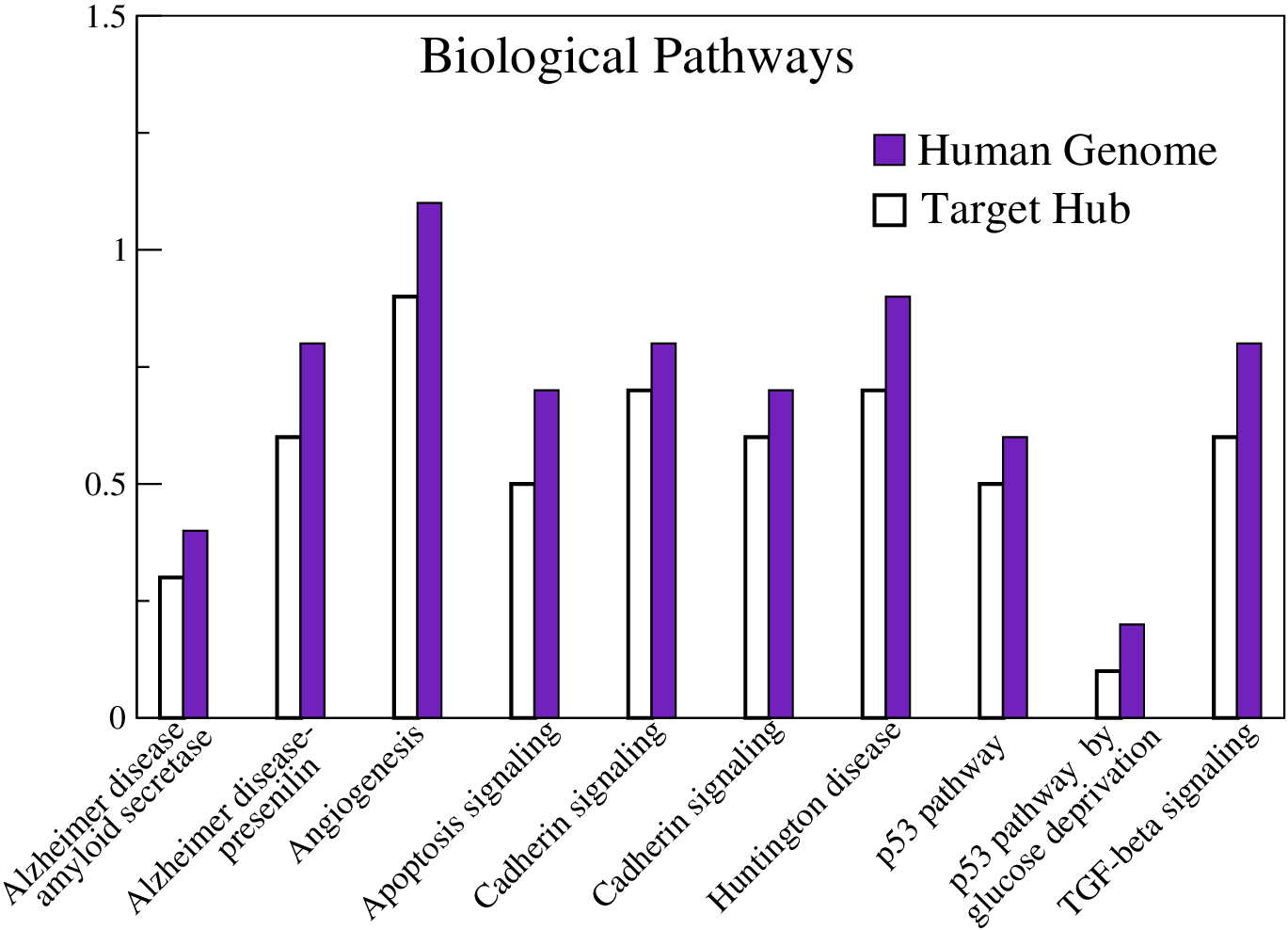}}
\textbf{Figure 2B:} Comparisons of various pathways in the human genome and among the genes in the target hubs. Representative pathways are shown where the genes in specific pathways are significantly enriched (filled bars) in comparison with that obtained in the human genome (open bars) using PANTHER.
\end{figure}

\subsection{Micro RNA co-target network}
To probe the detailed structure of the network, we create a co-target network of miRNA-miRNA as follows : if two miRNA co-target $w$ (non-zero) number of genes, we define that these miRNAs are linked. The weight of the link is $w$.  Otherwise, if $w=0$, no connection is made between these miRNAs. By doing so, we form an undirected weighted network of miRNAs. The corresponding adjacent matrix $C$ is symmetric having elements $C_{ij}$, where $C_{ij}$ is 
the number of genes co-targeted by miRNA $i$ and $j$. 
\begin{equation}
C=\left(
\begin{array}{ccccc}
C_{11}&C_{12}& \dots & \dots&C_{1~711}\cr
C_{21}&C_{22}& \dots & \dots&C_{2~711}\cr
&&\ddots&\cr&&&\ddots&\cr
C_{711~1}&C_{711~2}& \dots & \dots&C_{711~711}\cr
\end{array}
\right)
\end{equation}
 The diagonal elements $C_{ii}$, are  not well  defined. We take $C_{ii}=0$. Schematic description of this 
network is shown in the Figure 2C.

Few properties of this co-target network can be checked easily. First, it is a ``fully connected network'',  which does not have any sub-graph.  The distribution of weights $P_g(w)$ is shown in Figure 2D, which is again compared with  corresponding random graph. The result  indicates that  the connections are not random and the network 
is highly organized. Again we find that there is a huge difference between the maximum wight 
$(C_{ij})_{max}= 1253$ and the minimum $(C_{ij})_{min}= 1$, which leads to a conjecture that  
most links might  be irrelevant and can be erased. Next, we describe how to get an 
optimal set of miRNAs, by erasing  irrelevant links,  which could still describe all essential 
features of the co-target network.  

We define that the weak links are those whose weights are  smaller than a pre-decided cutoff $q$, and 
erase them. (For example when $q=10$, links between $(1, 6)$ and $(710, 3)$ in Figure 2C are to be erased.)
The network do not remain fully connected anymore and  breaks up into smaller 
disconnected sub-graphs.  Let the number of these disconnected sub graphs be $N_q$, which explicitly 
depends on $q$.   The largest among these subgraphs, named as G, is the \textit{important}  sub-graph. 
Clearly size of G  decreases with increase of $q$. For very large $q$, G has  only fewer miRNAs; that 
might simply the study of this network at the cost that some of the functionalities which are lost.   
Thus we need to optimize $q$ such that all irrelevant links are erased, keeping  only the optimal 
set of miRNAs in G  which are functionally relevent.  Note, that the cut-off $q$  redefines a  
new adjacency matrix
\begin{equation}
C_{ij}^q = \left\{ \begin{array}{cc} 0 & {\rm if}~ C_{ij} <q \cr 1&  {\rm otherwise} \end{array}\right. .
\nonumber
\end{equation}
Thus, the number of subgraphs $N_q$ is the number of diagonal blocks of $C^q$, which can be found by block-diagonalizing the matrix $C^q$. Since it is a fully connected graph, initially $N_1=1$ and then 
$N_q$ increases with q. Figure 2D shows a plot of $N_q$ versus $q$.

It is evident from Figure 2D that there are three regimes, (a) $q < q^*$, 
(b) $q \sim q^*$ and (c) $q> q^*$. Almost  all the connections are present in 
regime (a).  In this regime  the connections are quite \textit{stiff} and the network does 
not break up until a threshold $q=q^*$ is reached.
By increasing $q$ one goes to  regime (b) where  the irrelevant connections are 
already erased and only the relevant connections are present. Thus any little change in $q$ 
changes $N_q$ substantially. The relevant miRNAs, which cooperatively regulate the 
expressions of the genes to form the co-target network are probably found in regime (b). 
In regime (c),  $q$ is very large  to erase even the relevant  connections. Only connections 
which are present in this regime (c) are possibly due to chemical similarity, $i.e$, the 
miRNAs form very small subgroups and {\it seed} sequences (position $2$ to $7$ in the mature miRNA)
  of miRNAs within a subgroup 
are identical. This proposition has been verified as follows: we  take $q=600$ and 
find all the disconnected sub-graphs which have two or more miRNAs. The seed sequence of 
miRNAs in every sub-group has been checked. We find that the set of miRNAs in  each subgroup 
(except only one of 63 subgroups) have identical seed region 
[see supplementary table S2].

\begin{figure}
\centerline{\includegraphics[width=8cm]{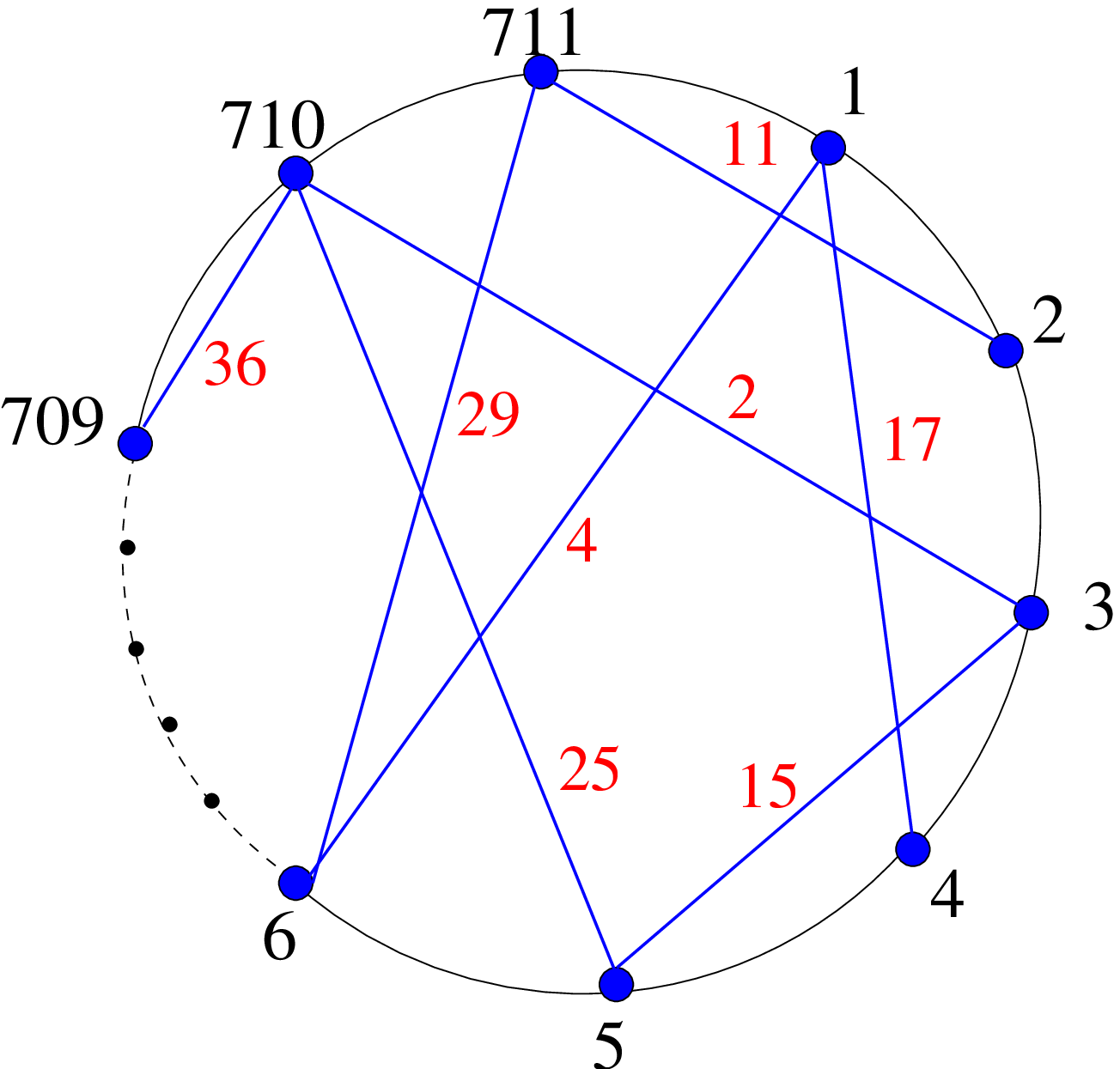}}
\textbf{Figure 2C:} Representation of the miRNA network. Here miRNAs (1 to 711) are shown in blue circles. In this example $C_{14}=17=C_{41}$ and $C_{23}=0= C_{3,2}$, which says that miRNA 1 and 4 co-target 17 genes whereas miRNA  2 and 3  do not have any common target. 
\end{figure}

Note that  the continuous break-up 
transition is a distinct property of miRNA co-target network, it does not 
occur for a random graph as shown in  Figure 2D (blue line).
To emphasize  this point further let us construct a 
random undirected graph $R$ which has total weight $W$ is same as that of $C$, 
 $i.e.$, 
 \begin{equation} 
  W= \sum_{i=1}^N \sum_{j=i+1}^M C_{ij}= 10354685=  \sum_{i=1}^N \sum_{j=i+1}^M R_{ij}
\end{equation}
Effectively, we need to choose $M(M-1)/2=252050$ 
random integers, one for each  upper diagonal element 
of $C$ ( total $M(M-1)/2$)  so that their sum is  $W$.  
Since the random matrix $R$ has to be symmetric ($R_{ij}= R_{ji}$), 
the lower diagonal elements can be calculated by substituting $R_{ij}$ by $R_{ji}$. 
In a similar way, by varying the cut-off $q$ we 
find $N_q$  for $R$, which is shown in Figure 2D (as blue dashed line).  
Note that the break up of the random network  into subgraphs 
in this case occurs discontinuously, unlike the 
miRNA co-target network. This result indicates that miRNA co-target 
network is structurally rich. Details topological structure of the 
network are now being  studied.

  The analysis reveals that the optimally connected graph (the sub-graph having maximum number 
of relevant miRNA nodes) which still posses the properties of the network can be found if we 
work at $q = q^*$. To find  $q^*$,  we differentiate $q$ numerically with respect to 
$q$.  The maximum of $\frac{d}{dq} N_q$  (as shown in the Figure 2D) corresponds to $q^*=103$.
In this case, at $q^*=103$, $N_q=166$ and the largest connected sub graph $\s$ is found to 
contain only $479$ nodes (for full list of miRNAs in 166 sub graphs see the supplementary table S3). 
Thus, these nodes or miRNAs form the optimal set of miRNA which co-regulate the gene expressions. 
To determined how miRNAs are organized within the sub graph $\s$, we increase the cutoff $q$ beyond $q^*$ and follow how $\s$ as breaks up into smaller clusters of miRNAs.  \textit{Clusters} in this context are defined as the 
subgraph which has more than one miRNAs. It is expected from Figure 2D that number of such clusters would not change much once we reach region (c) where  $\frac{d}{dq} N_q \simeq 0$ which starts approximately at  $q \approx 160$.  Thus, we take $q=160$ and collect the \textit{clusters }(the sub graphs of $\g$ which have two or more miRNAs). There are in total 70 such clusters containing 330 miRNAs. Thus it is reasonable to assume that the miRNAs within $\s$ are organized as 70 clusters, 169 independent miRNAs and interactions between them. These clusters form the basic units of the miRNA interaction network. Figure 3  shows the interaction network of miRNA of $70$ clusters. In the present study, we clustered the miRNAs on the basis of their ability to co-target the same genes, however, such co-targeting could be due to similarity in the seed sequences. We checked that 18 out of $70$ clusters might be due to seed similarity. About $50\%$ of the miRNA in human are organized within 10Kb of genomic sequence. Whether organizationally clustered miRNAs are under same regulatory sequence and expressed together remains unknown. Eleven out of $70$ clusters might be due to the genomic organization(Supplementary table S4).


We extended our  analysis to study  miRNA-miRNA interaction networks in Caenorhabditis elegans 
with $136$ miRNAs and their $20366$ targets (http://microrna.sanger.ac.uk/cgi-bin/targets/v5/genome.pl). The 
optimal value of $q$   in this  network is $q^*=23$. The largest connected sub graph at $q=q^*$ contains 
$98$ miRNAs. Most surprisingly, we find that both the species show universal features near the 
breakdown transition, similar to that shown for human in Figure 2D (data not shown). Reasons for 
such similarity in break down of miRNA-miRNA interaction network require further studies. 

It is also interesting to note that the clusters of miRNAs form a interaction network through their 
common targets (shown in Figure 3 and discussed in  section 4.4). Biological significance of this 
network in under investigation. 
  
\begin{figure}
\centerline{\includegraphics[width=8cm]{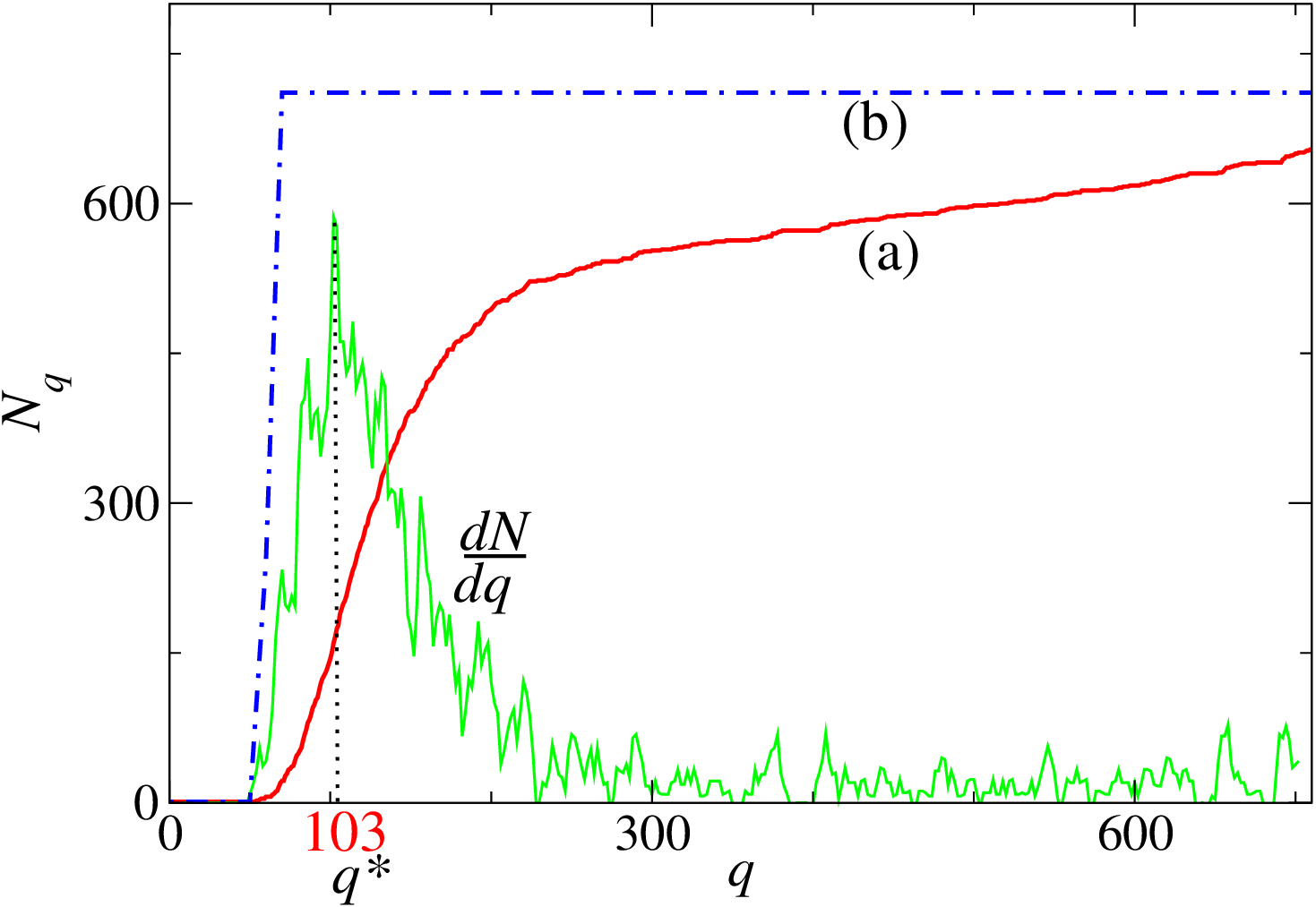}}
\textbf{Figure 2D:}  (a)  Number of  disconnected sub-graphs  $N_q$ when 
links  with weight  $<q$ are  erased. (b) corresponds to the same, for a 
random network. Plot of $\frac{d}{dq} N_q$  (which is scaled by a 
factor $20$ for better visibility)  shows a peak at $q^*=103$. 
\end{figure}


\subsection{Co-targeted genes in miRNA clusters may involve in specific pathways}
We need to know the biological relevance of the miRNA clusters better as they form the basic units of biological interactions for the regulation of the targets. The smallest cluster in subgraph $\g$, of course, contains two miRNAs, while the largest cluster contains $47$ miRNAs (Supplementary table S4). 
To explore whether the miRNAs in clusters target genes in specific pathways 
we determine if the co-targeted genes in a miRNA cluster is involved in 
any specific pathway. To do so, we determine the common targets of the miRNAs in a cluster taking two miRNA at a time, and then asked whether the co-targeted genes are enriched  for a particular pathway(s) compared to the targets which are not co-targeted by the miRNA in that cluster. For example, there are three miRNAs in cluster no. $25$ (supplementary table S4), namely miR-603, miR-521 and miR-523, which do not have common seed sequence and in combination [miR-603 and miR-521 ($n_1$)
and miR-521 and miR-523 ($n_2$) and mir-523 and miR-603  ($n_3$)] 
target 
$449$ genes [($n_1+ n_2 + n_3$)]. These three miRNAs together target $2309$ genes. 
The number of genes not targeted by the miRNAs in this cluster is $1860$. 
Using the PANTHER, we observed that common targeted genes ($449$) are significantly ($p \le 0.05$) enriched in 
axon guidance mediated by semamorphins (P00007), cell cycle (P00013), DNA replication (P00017) and FGF signaling (P00021) pathways. This analysis further reveals that out of $70$ clusters, co-targeted genes in $47$ clusters  are significantly enriched at least one biological pathway (total biological pathways in human at PANTHER database is $153$). In several cases, the co-targeted genes are over represented in more than one pathway (supplementary table S4). Significant enhancement of  the genes co-targeted by the miRNAs in a cluster indicate that these clustered miRNAs regulate genes in  specific pathway.

\begin{figure}
\centerline{\includegraphics[width=8cm]{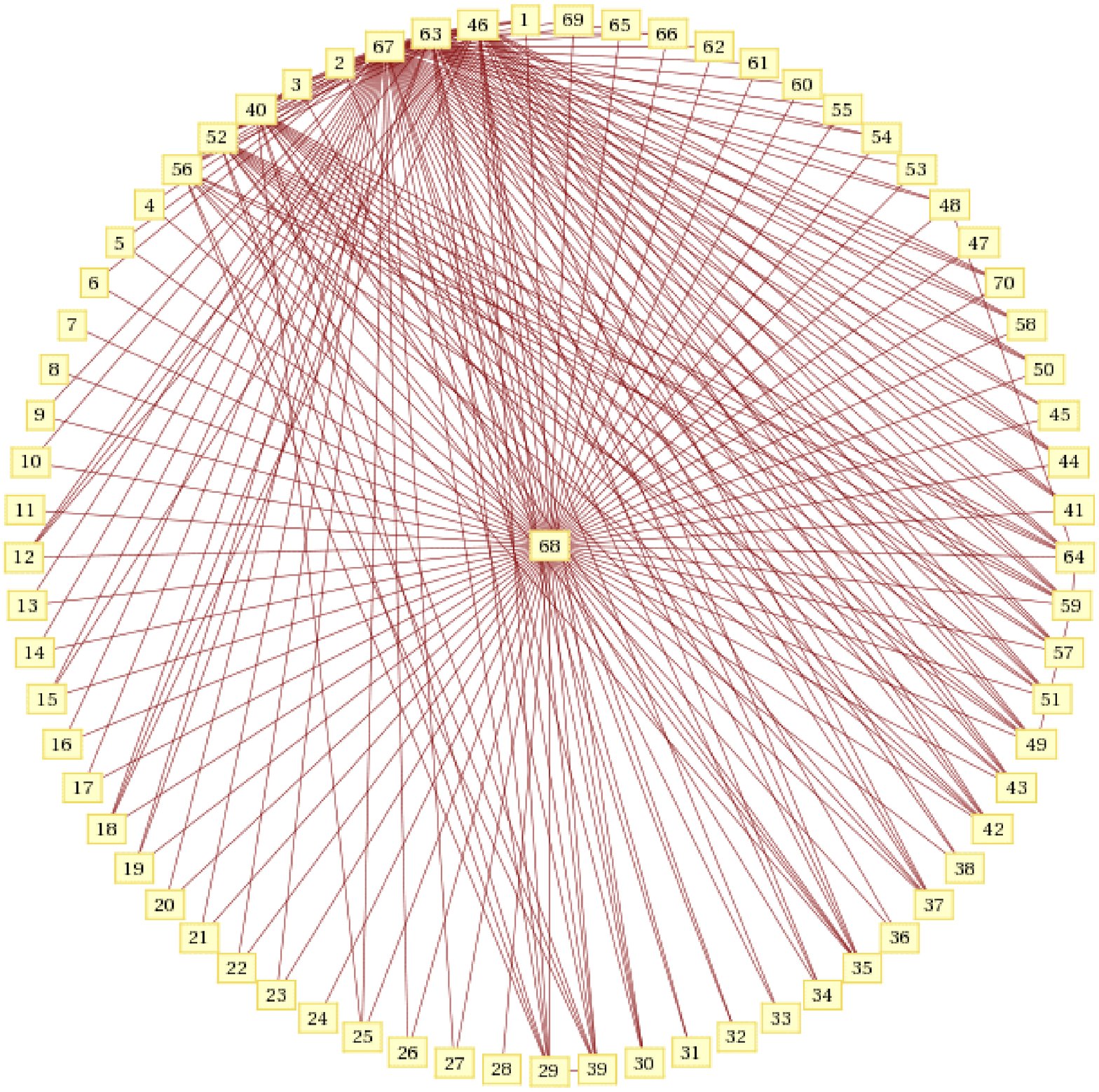}}
 \textbf{Figure 3:} Interaction network of $70$ clusters containing 330 miRNAs (
links with weight $< 2000$ are erased).  
\end{figure}

 \subsection{Involvement of clustered miRNAs in human diseases} 
Alterations of several miRNAs have been observed experimentally in about eighty diseases [25]. To probe whether the clustered miRNAs obtained above are associated with any disease, we retrieved data from published papers, where deregulated miRNAs in diseases are reported. We collected data for 15 diseases where deregulation of miRNAs are reported (Supplementary table S5, see supplementary text for the references). These diseases include autism (number of deregulated miRNA 31), schizophrenia (22), diabetes (glucose homeostasis) (19), ovarian cancer (54), AML (38), breast cancer (28), colon cancer (19), lung cancer (39), pancreas cancer (56), prostate cancer (48), stomach cancer (25), HNC (59), thyroid cancer (51), CLL (53) and glioblastoma (11) (Supplementary table S5). We note that total number of miRNAs studied in these reports are not available for all the cases, making it difficult to assign the miRNAs which are not changed in these diseases. Among the altered miRNAs, number of miRNAs in the largest cluster G (479 miRNA) with q=103, and in the 70 clusters (S) derived from the G with $q=160$ is shown in the TABLE I. It is evident from the table that for some diseases, the numbers of miRNAs in G are significantly higher than one expects from random distributions of the miRNAs in clusters. For example, in prostate cancer, 48 miRNAs are deregulated, out of which 41 miRNAs  belong to the cluster G (479 miRNA). The expected number of miRNA in this cluster is 32 ($479\times 48/711$), which is significantly lower from the observed value of 41 miRNA. It is interesting to note that when the largest cluster G was broken into 70 inter-connected clusters (S), the level of significances of the miRNAs among 70 clusters than the expected values (shown in TABLE I) are increased considerably. This trend further justified the breaking of the largest cluster/sub-graph G into smaller ones.

\begin{table*}
\caption{ Significant number of miRNAs  associated with any particular disease 
(known from published data) are present in G. Even more significantly they belong to the subgroup 
group S consisting of \textit{cluster forming} miRNAs.}
\begin{ruledtabular}
  \begin{tabular}{|c|c|c|c|} 
{\bf Disease}	& No of miRNAs &No of miRNA in $\s$& 
No of  miRNA in $\g$	\cr  
 & (from published data)&(expected)~[$p$-value]& (expected)~[$p$-value]	\cr \hline
Autism &		31&	26	(21)~~[0.050]&	20	(14)~~[0.043]\cr \hline
Schizophrenia&		22&	19	(15)~~[0.057]&	18	(10)~~[0.001]\cr \hline
Diabetes&		19&	13	(13)~~[0.922]&	9	(9)~~~~[0.933]\cr \hline
Overian Cancer& 		54&	41	(36)~~[0.180]&	32	(25)~~[0.058]\cr \hline
AML & 			38&	29	(26)~~[0.240]&	23	(18)~~[0.081]\cr \hline  
Breast Cancer&		28&	23	(19)~~[0.095]&	18	(13)~~[0.058]\cr \hline
Colon Cancer&		19&	13	(13)~~[0.922]&	10	(9)~~~[0.587]\cr \hline
Lung Cancer&		39&	28	(26)~~[0.556]&	20	(18)~~[0.542]\cr \hline
Pancreas Cancer&	56&	41	(38)~~[0.351]&	35	(26)~~[0.016]\cr \hline
Prostate Cancer&	48&	41	(32)~~[0.008]&	38	(22)~~[0.000]\cr \hline
Stomach Cancer&		25&	18	(17)~~[0.621]&	15	(12)~~[0.173]\cr \hline
 HNC &	59&	50	(40)~~[0.004]&	39	(27)~~[0.002]\cr \hline
Thyroid Cancer&		51&	40	(34)~~[0.092]&	32	(24)~~[0.019]\cr \hline
CLL  & 		53&	44	(36)~~[0.015]&	36	(25)~~[0.002]\cr \hline
Glioblastoma&		11&	9	(7)~~~~[0.307]&	9	(5)~~~~[0.019]\cr 
  \end{tabular}\\
AML: \begin{footnotesize}Acute Myeloid  Leukemia\end{footnotesize}, 
HNC : \begin{footnotesize}Head and Neck  Cancer\end{footnotesize}, 
CLL : \begin{footnotesize}Chronic Lymphocytic Leukemia  \end{footnotesize}
\end{ruledtabular}
\end{table*}

	The miRNAs deregulated in the diseases are distributed among the 70 clusters in S and compared with that was expected from random distributions.  For example, out of 28 miRNA deregulated in breast cancer, 18 miRNAs are distributed in 11 clusters (cluster \# 3, 16, 20, 35, 37, 51, 57, 63, 64, 67 and 68, as shown in  the supplementary table S4) among the 70 clusters. Among these 11 clusters, where 18 deregulated miRNAs are distributed, 8 clusters (cluster \# 3, 16, 20, 35, 37, 51, 57 and 64, see supplementary table S4 for the clusters) have significantly higher number of miRNA than that one expects from random distribution. Similarly, out of 22 miRNA deregulated in schizophrenia 18 miRNA in S are distributed in 11 clusters. Among these 11 clusters only 8 clusters have significantly higher number of miRNA than one expects.  Distributions of miRNAs in different clusters are significantly different from that expected from random distributions. The representative result is shown in the Figure 4 and the detailed result for all these 15 diseases are shown in the supplementary Figure SF1. This result shows that significant numbers of deregulated miRNAs in the diseases studied in this investigation belong to specific clusters. Only the clusters, which harbor significantly higher number of miRNA over the random distributions, are considered for further analysis (see next section).  Common targeted genes in different clusters are significantly enriched in different pathways as shown in the earlier section above.

Common targeted genes of the clusters 4, 55, 57 and 64 are enhanced in Huntington’s disease pathway (P00029). Cluster 4 consists of miR-423-3p and miR-24. This cluster 4 contains significantly more deregulated miRNA in schizophrenia, colon cancer, pancreatic cancer, prostate cancer, stomach cancer, head and neck cancer and thyroid cancer (supplementary Figure SF1) than expected from the random distributions. This result indicates that Huntington’s disease pathway might be deregulated in these diseases. To search further if there is any experimental proof for this proposition, we searched “Huntington disease and Schizophrenia” in “pubmed”. More than 300 references are 
found to contain both the terms. Evidence that schizophrenia-like symptoms in Huntington’s disease pedigree indicate that at molecular level the Huntington’s disease pathway and schizophrenia might have some overlap (Corrêa et al., 2006). In Huntington’s disease (HD), accelerated loss of striatal neurons by apoptotic death is the major cellular event (Gil and Rego, 2008), while in cancer cell death is prevented in general; apoptosis thus might play opposite role in HD and cancer. Epidemiological study indicates that the prevalence of cancer among HD patients are less than that observed among individuals without HD (Sorensen et al., 1999). Knock out of p53 gene, a tumor suppressor gene and mutated in $~50\%$ of the cancers, induces tumors in mice and decrease the life span of the mice. Recently, it has been shown that the expression of the mutant allele of HTT gene (expansion of CAG repeats beyond 36 at the exon1 causes HD) prolonged the life of p53 knocked out mice. This observation has been explained by the enhanced apoptosis induced by the mutant HTT, which is likely to prevent cancer (Ryan and Scrable, 2008). This result further indicates that Huntington disease pathway may be altered in cancer. Our result that specific cluster of miRNA target this pathway indicate that in cancer as well as schizophrenia this pathway might be altered.   

We  further tested whether specific clusters of miRNA are involved in specific disease. We collected the mRNA expression data from the published literature for autism (number of deregulated mRNA 795), schizophrenia (146) and diabetes (189) [supplementary table S6 and supplementary text]. Similar data for ovarian cancer (6219), AML (5440), breast cancer (6100), colon cancer (932), lung cancer (2747), pancreas cancer (6950), prostate cancer (4410), stomach cancer (589), HNC (5468), thyroid cancer (3092), CLL (1967) and glioblastoma (6528) was taken from ONCOMINE (http://www.oncomine.org/) and determined whether deregulated genes are enriched in specific pathway(s). Detailed result is shown in the supplementary table S6. It is to be noted that only a subset of these deregulated genes are likely to be deregulated by miRNAs. We compared the pathways that are enriched for the deregulated genes 
in comparison to that obtained among genes coded by human genom.
in microarray data with that are obtained for deregulated miRNAs in these diseases and significantly enriched in the clusters as described above.  As mentioned above, in breast cancer, 28 miRNAs are deregulated; 18 miRNA are distributed in 11 clusters among the 70 clusters of which 8 clusters harbors miRNAs significantly over the random distribution. Common targeted genes in these 8 clusters are over represented in 12 unique biological pathways (total 13 pathways, supplementary table S7). We then compared the pathways that are enriched for the deregulated genes and observed that 10 unique pathways are common.  Common pathways that are enriched in the specific clusters of miRNAs and deregulated genes in diseases further indicate that the specific clusters of miRNA are likely to target specific pathway(s) and involve in diseases.

Significance of ``co-targeting the genes'' by miRNA is not clear and requires further investigations. Limited experimental evidences indicate that miRNA influences the expressions of the targets moderately (Lim et al., 2005, Selbach et al., 2008) and proposed to be involved in “fine-tuning” of gene expressions (Flynt and Lai, 2008). Thus, in combination, the level of expression of target genes of miRNA might reduce further the gene expression compared to that could be obtained with a single miRNA. Combinational regulation of target genes by miRNAs might be complicated by the location of the target sites at $3^\prime$ UTR as well as the tissue specific expressions of the miRNAs. There are overlapping or very closely spaced miRNA target sites for a particular target gene. Such target gene might not have additional effects of the second miRNA due to physical hindrance of the target site by binding of the other miRNA at the near by or overlapping site. The same gene may be targeted by different miRNA, when the miRNA expression is regulated at the tissue level.

\begin{figure}
\centerline{\includegraphics[width=8cm]{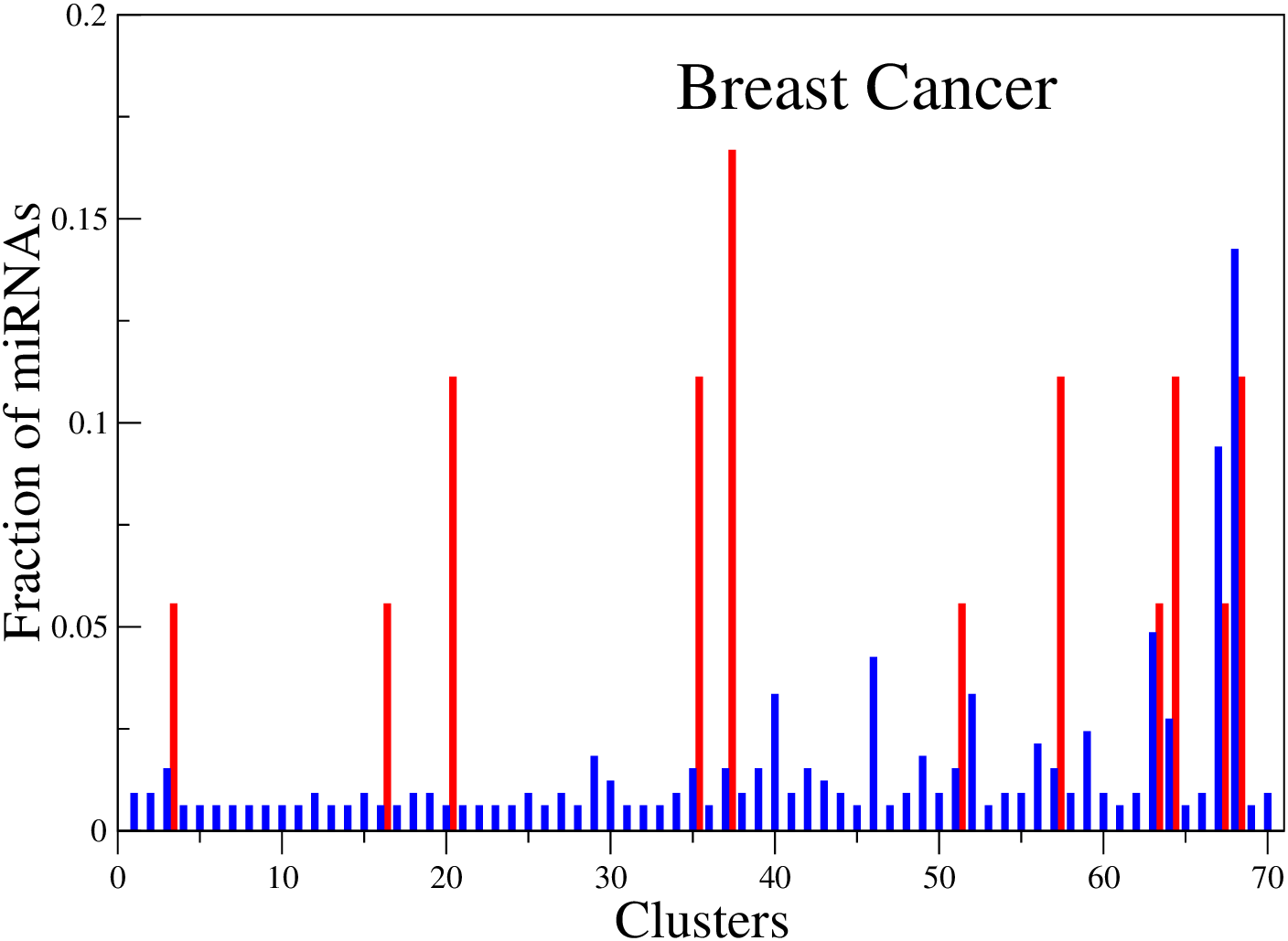}}
\textbf{Figure 4:} Distribution of deregulated miRNAs in breast cancer (29). $19$ out of 
$29$ miRNAs are distributed in $11$ clusters. Significant increase is found only in 
$8$ cluster ( \# $ 3, 16, 20, 35, 37, 51, 57, 64$).
\end{figure}

In our analysis seventy clusters of miRNA contains all together 330 miRNA and each cluster contains has than 2 miRNA. Out of these 70 clusters, 18 clusters could arise due to the seed similarities and 11 clusters could arise for genomic organization (co-localized). Among these clusters, 4 clusters are common for the seed similarity and co-localization (Supplementary table S4). It is interesting to note that out of 18 clusters, which could arise due to seed sequence similarity, only 5 clusters are associated with pathways, which are enriched with co-targeted genes. Two clusters (\#26 and \#31) that could arise from seed sequence similarity and co localization, also associated with co-targeted genes enriched in specific pathways. This result indicates that in the majority of cases, clustered miRNA did not arise due to seed sequence similarity.  Co-targeted genes in the miRNA cluster and enriched in specific pathways for 5 clusters could arise from co-localization of the miRNAs in the genomic regions. These miRNAs, which are co localized in the same chromosomal region (with in 10Kb upstream and down stream of the pre-miRNA) might be regulated by the same promoter regions and co-expressed together. In a recent study, miRNAs hsa-mir-92a-1, hsa-mir-20a, hsa-mir-18a, hsa-mir-17, hsa-mir-19b-1 and hsa-mir-19a, which are co localized on chromosome 13 within 10kb, have been shown to act in a combinatorial fashion. The authors explained that in combination, the miRNA down regulated to the measurable extent, while individual miRNA was unable to down regulate the target sufficiently [21]. This result also support the notion that miRNAs in combination lowers the expressions than an individual miRNA. In the majority (47/70) of the miRNA clusters, co-targeted genes are enriched in at least one pathway. Out of 153 pathways described in the PANTHER (total number of genes 25431, total number pathway hit in 7151, thus one gene is involved in more than one pathways), 105 distinct pathways are enriched with the co-targeted genes in 47 clusters. Among these 47 clusters, 28 clusters contain significantly excess deregulated miRNAs in diseases. Co-targeted genes in these 28 clusters are enriched in 72 distinct pathways. Same pathway appeared in multiple times in different diseases, altogether 72 pathways appeared 359 times in diseases. For example, Huntington’s disease pathway (P00029) appears 19 times in 15 diseases indicating the importance of this pathway in several diseases. The miRNA clusters where co-targeted genes are enriched in this particular pathway and also harbors significantly higher deregulated miRNA in diseases are 4 (423-3p, miR-24), 33 (miR-149, miR-892b), 52 (miR-508-5p, miR-516a-3p, miR-198, miR-520a-5p, miR-517*, miR-525-5p, miR-516b, miR-518c*, miR-518e*, miR-518d-5p, miR-518f*), 55 (miR-373*, miR-616*, miR-888), 64 (miR-331-3p, miR-146b-3p, miR-18b*, miR-18a*, miR-324-5p, miR-874, miR-324-3p, miR-10a, miR-10b). These miRNAs are likely to target 172 genes in Huntington disease pathway and participates in the diseases discussed in this manuscript. It is interesting to mention, although significance if any remains unknown, that Huntington’s disease pathway was also over represented among the genes at the target hub as described in earlier section. Experimental verification to substantiate of the contention that miRNAs in the clusters together alter this pathway is necessary to confirm the prediction.

\section{Conclusion} 
In conclusion, we constructed a miRNA-miRNA weighted interaction network using 711 miRNAs and their predicted targets. A novel method was applied to break up the network into smaller sub graphs (clusters). We further extended our studies to show that specific clusters of miRNA might be involved in specific pathways that are known/predicted in several diseases. We propose that instead of a single miRNA, a group of miRNA together target genes in specific pathway and the interactions of these clusters are likely to be involved in diseases.   

\section{Method}
 \subsection{Data mining} 
	All the miRNA and the predicted targets are downloaded from miRBase (http://microrna.sanger.ac.uk/, version 10). In version 10 of this database, there are 711 miRNAs coded by 533 genomic loci. This is because, double stranded miRNA produced by DICER and other protein is separated by helicase, and in some cases both the strands can act as the mature single stranded miRNA. In general, the expression of one of the strand is more than the others.  The strand, which expresses less, is denoted by mir*.  Predicted numbers of transcript (mRNA) that are targeted by this 711 miRNA is 34525. It is to be noted that total number of human genes in PANTHER is 25431. Thus, it is likely that the some of the targets of miRNA in the Sanger database may actually be the isoform of the genes. All the predicted targets of miRNA in miRBase are identified with the ENSEMBLE ID (http://www.ensembl.org/). For analysis with different data bases, the IDs are converted into NCBI gene IDs by using the site Clone/Gene Id Converter (http://idconverter.bioinfo.cnio.es/) and Ensemble Genome Browser (http://www.ensembl.org/index.html).  

Genes which are down regulated or up regulated in various cancers are downloaded from ONCOMINE (http://www.oncomine.org/). For other diseases, we collected the deregulated genes from various published data. ONCOMINE provides the gene names/symbols. These gene names are converted into the NCBI gene Id  (NM numbers) by using the site Clone/Gene Id Converter (http://idconverter.bioinfo.cnio.es/) and by using various web sites as mentioned above. The deregulated miRNAs in different diseases are also collected from the different published data. Full references are provided in the supplementary text. 

 \subsection{Formation of  network}

We use in house perl scripts to convert the dataset containing $711$ miRNAs and their targets 
into  adjacency matrix
$A_{ij}$ (see text) by giving arbitrary, but unique, identification numbers 
to genes as $i=1,2\dots N$ and miRNAs as $j=1,2\dots M$ respectively. 
$A_{ij}$ may take value $1$ if  gene $i$ contains a predicted recognition site 
for miRNA $j$ in the $3^\prime$ UTR  or otherwise $A_{ij}=0$. Corresponding network is 
shown in Figure 1,  
where all the genes (miRNAs) are represented as  red(blue) circles and a line 
(joining blue circle $j$ red circle$i$) is drawn  if  $A_{ij}=1$.  

     Further, matrix $A$ is used for constructing a miRNA co-target network.  Two 
miRNAs $j$ and $k$  are said to be connected  if they have at least one common 
target. Weight of the connection or link is  defined to be $C_{jk}$, which is 
simply the  number of common targets of miRNA $j$ and $k$. Matrix $C$  has dimesion $711\times 711$, which  can be constructed 
from $A$  as $C= A^TA$, where transpose of $A$ is defined as $(A^T)_{ji}= A_{ij}$.
Or in otherwords, 
\begin{equation}    
C_{jk}=  \sum_i A_{ij}A_{ik}.
\end{equation}

  \subsection{Clustering of miRNAs in miRNA co-target network}
   Since the miRNA co-target network is fully connected with large number of 
links having very small weights, we would like to  get rid of these 
links  and find  sub-graph which contains only relevent miRNAs. This can be done 
by erasing  the  weak links, say the links which has weight less than a 
cutoff $q$. The network, then breaks up  into several disconnected sub-graphs, 
in total $N_q$. Thus, $N_q$ is just the number of diagonal blocks of matrix $C$, which 
was calculated using an in house {\bf C} program. $N_q$ changes marginally as 
$q$ is increased and then a  rapid change occurs at $q=q^*$.  It is assumed 
that the network breaks rapidly as  the relevent (or most significant) links 
are removed.  To find  $q^*$, we differentiate $N_q$  with respect to $q$ 
numerically and find the peak position, which corresponds to  the $q$ value
 where the change is maximum.  Now we fix $q=q^*$ and find the 
largest sub-graph $\s$ among $N_{q^*}$ subgraphs.  G is considered to be the 
relevant.
 To know, how the miRNAs are arranged 
within $\s$  we further increase $q$  to $160$ (see text for the explanation on 
why we choose $q=160$) and  collect the subgraphs of 
$\s$.  The sub-graph $\s$ containing more than   one miRNA are  called {\it clusters}.  
  \subsection{Network of miRNA clusters} To get a network of miRNA clusters, first  we get the 
weight $W_{mn}$ of the link between cluster $m$ and $n$. Let cluster $m$ ($n$) has $N_m$ 
( $N_n$) miRNAs. We add  the weights(no  of co-targets) of every pair miRNA, fromed by taking 
one miRNA from cluster $m$ and the other from cluster $n$, and define the sum to be the weight of the link.
 Thus, $W$   forms a $(70\times 70)$ matrix with elements
\begin{equation}
W_{mn}= \sum_{i\in m} \sum_{j\in n}  C_{ij}.
\end{equation}
The diagonal elements $W_{nn}$ are non-zero and it indicates the strength of the cluster which is the  total number of pair-wise co-targets  of the cluster $n$.  There is no natural cutoff 
on the weights as they show a scale free distribution (data not shown). We take an arbitrary cutoff $2000$ to 
draw the network, $i.e$, a link is drawn between cluster $m$ and $n$ only if $W_{mn}>2000$.

\subsection{Co-targeted genes in miRNA clusters} 
 For the identification of target genes in a specific cluster, we first find the co-targeted genes pair wise in a cluster. For example, if there are three miRNA like miR-1, miR-2 and miR-3 (hypothetical) are in a cluster, 
the predicted common targets of the miRNAs are those, which are common for miR-1 and miR-2 ($n_1$), miR-2 and miR-3 ($n_2$), miR-1 and miR-3 ($n_3$). The pathway(s) where these common targets ($n=n_1+n_2+n_3$) belong are obtained from the PANTHER. This result was compared with that obtained with those targets, which are not common ($m-n$ in total, where $m$ is the total predicted targets of miR-1, miR-2 and miR-3). 
 
  \subsection{Classification of the genes in target hub, co-targeted by miRNA in clusters, deregulated in diseases using PANTHER and enhancement analysis }
	In PANTHER (Protein Analysis Through Evolutionary Relationships), genes are classified on the basis of their functions using published experimental observations and evolutionary relationships; in the absence of direct experimental evidence. Proteins are classified by expert biologists and categorized by molecular function and biological process and biological pathways (http://www.pantherdb.org/). 
In miRNA target search, targets are identified with the Ensemble IDs that contains several isoforms of 
the genes. We omitted these isoforms as well as other Ensemble IDs for which we are unable to retrieve the 
Gene symbols for PANTHER analysis. 
To identify, whether genes in target hub, co-targeted by miRNA in clusters and deregulated in diseases when classified on the basis of molecular functions, biological processes and pathways are over represented (enhanced) in particular category in comparison with that obtained in the human genome we analyzed the genes using the PANTHER. If the fraction of genes in the test samples are significantly enriched in a particular category over that obtained in the human genome, we consider the specific category is involved/associated with the test samples. For example, if the co-targeted genes in a particular cluster are significantly over represented in Wnt signaling pathway (PANTHER ID P00057), we consider that this cluster was associated with this pathway.
	
Deregulated expressions of genes observed in several diseases  discussed in this 
manuscript which are taken from  published literature are analyzed using PANRHER.
Significant ($p$ values less than $0.05$) increase in genes in  specific pathways in comparison with that of the human genom are identified. 

  \subsection{Correlation of miRNAs in clusters and diseases}
From the published literature we collected $15$ diseases where the deregulation of 
several miRNAs are reported. We then ask: what fraction these reported miRNAs, say 
$n$ in total, are distributed among the $70$ different clusters (obtained  at $q=160$, see text for details). This fraction is then compared with the random distribution of 
$n$ miRNAs  among $70$ clusters.  Those clusters having miRNAs significantly larger   (with $p$ values less than $0.05$)  than that of a random distribution  are considered  relevant  for this particular disease. Significantly enriched genes in pathways, identified from common miRNA targeted genes in  these relevant clusters, are compared with the pathways of significantly enriched genes observed in deregulated mRNAs 
(from micro-array data).
  \subsection{Comparisons of the pathways between those observed in common targeted genes in clusters and that of in  deregulated mRNA in diseases}

From ONCOMINE and published literature we collected the deregulated genes in 15 diseases and analyzed using PANTHER. Significantly altered pathways are identified. These are then compared with the pathways obtained   in common targeted genes in clusters.

  \subsection{Statistical analysis}
Calculation of $p$ value can be best described with an example. If 
$n=11$ is the number of miRNA present in a cluster among total $N=330$ miRNA, and for 
a particular disease there are only $m=2$ miRNAs in that cluster among total $M=30$ miRNAs,
then  $\chi^2 = (mN/M -n)^2 [ 1/n + 1/(N-n)]$, which is $11.38$ in this example. $p$ 
value is the probability that $x>\chi^2$  in {\it chi-square} distribution $Q(x)$.
Thus, $p= 1-\int_0^{\chi^2} Q(x) dx$ which can be readily integrated  using Mathematica
or  by reading it from  a Table. In this  example $p=0.00074$. Note that $p<.050$ is  
equivalent to  $\chi^2>3.841$.

\section{Author Contributions}
PKM and NPB conceptualized the problems, PKM, SUSHMITA, MS and SM carried out the computational analysis, PKM, SUSHMITA and NPB analyzed the data and wrote the 
manuscript.

\section{Supplementary Matterial}
Kindly E-mail PKM (\textbf{pk.mohanty@saha.ac.in}) to get the  Supplementary Matterial.

\subsection{ Supplementary tables}

\textbf{Table S1}: List of the genes that targeted by more than 20 miRNAs (target hub) [S1A].  Genes at target hub are classified using PANTHER according to molecular functions, biological processes and biological pathways and compared with that obtained with the human genome (S1 B, S1C and S1D). 

\textbf{Table S2:} List of the miRNAs that have same seed regions (position $2$ to $7$ in the mature miRNA)  nucleotides of the the mature miRNAs.

\textbf{Table S3: } List of miRNAs in 166 sub graphs which are found at $q^*=103$.

\textbf{Table S4:} List of 70 clusters and the miRNAs belonging to each cluster, seed sequence similarity, genomic organization (only those which are clustered together within 10kb genomic region are only noted). In a cluster, the common targets for each pair of miRNA is determined and added. Biological pathways for the common targets in comparisons with the total targets that are not common are compared. Only the pathways that are significantly (less than equal to 0.05) enriched in the targets are shown. 

\textbf{Table S5:} Experimentally observed deregulated miRNAs in 15 different diseases are shown. For references of the sources of these miRNA please see the supplementary text.

\textbf{Table S6: }Analysis of deregulated genes in 15 different diseases is shown. For references of the sources of the experimentally determined deregulated genes please see the supplementary text. 

\textbf{Table S7: } Significantly enhanced genes in pathways for altered mRNA is listed for 15 different diseases. 
Those pathways which are common to the pathways  predicted from the miRNA clusters are marked (in red).

\subsection{ Supplementary Figures}

\textbf{SF1: } Distributions of various experimentally determined deregulated miRNAs among 70 clusters of miRNAs 

\subsection{ Supplementary Text}
References of the literature from where we have collated the deregulated miRNA and genes.


\begin{thebibliography}{99}


\bibitem{1.}Liu J (2008) Control of protein synthesis and mRNA degradation by microRNAs. Curr Opin Cell Biol. 20: 214-221

\bibitem{2.} Place RF, Li LC, Pookot D, Noonan EJ, Dahiya R (2008) MicroRNA-373 induces expression of genes with complementary promoter sequences, Proc Natl Acad Sci U S A. 105, 1608-1613

\bibitem{3.}Tay Y, Zhang J, Thomson AM, Lim B, Rigoutsos I (2008) MicroRNAs to Nanog, Oct4 and Sox2 coding regions modulate embryonic stem cell differentiation, Nature, 455: 1124-1128

\bibitem{4.} Ørom UA, Nielsen FC, Lund AH (2008) MicroRNA-10a binds the $5^\prime$UTR of ribosomal protein mRNAs and enhances their translation, Mol Cell, 30, 460-471.

\bibitem{5.} Lewis BP, Burge CB, Bartel DP (2005) Conserved seed pairing, often flanked by adenosines, indicates that thousands of human genes are microRNA targets.     Cell, 120:15-20.

\bibitem{6.} Miranda KC, Huynh T, Tay Y, Ang YS, Tam WL, Thomson AM, Lim B, Rigoutsos I (2006) A pattern-based method for the identification of MicroRNA binding sites and their corresponding heteroduplexes. Cell, 126:1203-1217

\bibitem{7.} Grimson A, Farh KK, Johnston WK, Garrett-Engele P, Lim LP, Bartel DP (2007) MicroRNA targeting specificity in mammals: determinants beyond seed pairing. 
            Mol Cell. 27, 91-105

\bibitem{8.} Volinia S, Calin GA, Liu CG, Ambs S, Cimmino A, Petrocca F, Visone R, Iorio M, Roldo C, Ferracin M, Prueitt RL, Yanaihara N, Lanza G, Scarpa A, Vecchione A, Negrini M, Harris CC, Croce CM (2006) A microRNA expression signature of human solid tumors defines cancer gene targets. Proc Natl Acad Sci U S A.103: 2257-2261.

\bibitem{9.}Dalmay T (2008) MicroRNAs and cancer, J Intern Med 263: 366-375

\bibitem{10.} Thum T, Catalucci D, Bauersachs J (2008) MicroRNAs: novel regulators in cardiac development and disease. Cardiovasc Res. 79, 562-570

\bibitem{11.}Zhang C (2008) MicroRNomics: a newly emerging approach for disease biology. Physiol Genomics, 33:139-147

\bibitem{12.} Sonkoly E, Ståhle M, Pivarcsi A (2008) MicroRNAs and immunity: novel players in the regulation of normal immune function and inflammation. Semin Cancer Biol. 18:131-40

\bibitem{13.} Blakaj A, Lin H (2008) Piecing together the mosaic of early mammalian development through microRNAs. 283:9505-9508 

\bibitem{14.} Stefani G, Slack FJ (2008) Small non-coding RNAs in animal development.
            Nat Rev Mol Cell Biol. 9:219-23015.
\bibitem{15.} Cheng AM, Byrom MW, Shelton J, Ford LP (2005) Antisense inhibition of human miRNAs and indications for an involvement of miRNA in cell growth and apoptosis. Nucleic Acids Res. 33:1290-1297.

\bibitem{16.} Flynt AS, Lai EC (2008) Biological principles of microRNA-mediated regulation: shared themes amid diversity. Nat Rev Genet., 9, 831-842

\bibitem{17.} Miska EA, Alvarez-Saavedra E, Abbott AL, Lau NC, Hellman AB, McGonagle SM, Bartel DP, Ambros VR, Horvitz HR (2007) Most Caenorhabditis elegans microRNAs are individually not essential for development or viability. PLoS Genet. 3, e215

\bibitem{18.}Cui Q, Yu Z, Purisima EO, Wang E (2006) Principles of microRNA regulation of a human cellular signaling network. Mol Syst Biol. 2:46. 

\bibitem{19.} Zhou Y, Ferguson J, Chang JT, Kluger Y (2007) Inter- and intra-combinatorial regulation by transcription factors and microRNAs. BMC Genomics.8: 396

\bibitem{20.} Shalgi R, Lieber D, Oren M, Pilpel Y (2007) Global and local architecture of the mammalian microRNA-transcription factor regulatory network, PLoS Comput Biol 3:e131

\bibitem{21.} Ivanovska I, Cleary MA (2008) Combinatorial microRNAs: Working together to make a difference, Cell Cycle, 7: 3137-3142.

\bibitem{22.} Gu P, Reid JG, Gao X, Shaw CA, Creighton C, Tran PL, Zhou X, Drabek RB, Steffen DL, Hoang DM, Weiss MK, Naghavi AO, El-daye J, Khan MF, Legge GB, Wheeler DA, Gibbs RA, Miller JN, Cooney AJ, Gunaratne PH (2008) Novel microRNA candidates and miRNA-mRNA pairs in embryonic stem (ES) cells, PLoS ONE, 3(7), e2548.

\bibitem{23.}Lim LP, Lau NC, Garrett-Engele P, Grimson A, Schelter JM, Castle J, Bartel DP, Linsley PS, Johnson JM (2005) Microarray analysis shows that some microRNAs downregulate large numbers of target mRNAs, Nature, 433:769-773.

\bibitem{24.} Selbach M, Schwanhäusser B, Thierfelder N, Fang Z, Khanin R, Rajewsky N (2008) Widespread changes in protein synthesis induced by microRNAs, Nature, 455:58-63

\bibitem{25.} Lu M, Zhang Q, Deng M, Miao J, Guo Y, Gao W, Cui Q (2008) An analysis of human microRNA and disease associations, PLoS ONE. 3:e3420
\end{thebibliography}
\end{document}